\documentclass[acmsmall]{acmart}
\AtBeginDocument{%
  \providecommand\BibTeX{{%
    \normalfont B\kern-0.5em{\scshape i\kern-0.25em b}\kern-0.8em\TeX}}}
\setcopyright{rightsretained}
\copyrightyear{2021}
\acmYear{2021}
\acmDOI{1}
\acmJournal{TECS}
\acmVolume{1}
\acmNumber{1}
\acmArticle{1}
\acmMonth{1}
\RequirePackage[T1]{fontenc}

\usepackage{amsthm}
\usepackage{verbatim}
\usepackage{paralist}
\providecommand\doi[1]{\href{https://doi.org/#1}{\url{#1}}}

\usepackage[bbsets,Dfprime]{math}
\usepackage[prefixflatinterpret,bracketmodalinterpret,modernsign,substopindex,shortmquant,mquantifiertype,mconnectiveformal,bracketinterpret,fixformat,setfixinterpret,modifopindex,seqarrow,seqoptional,sidenotecalculus,abbrseqcontext,shortterms,nosigmaterms,novarterms]{logic}
\usepackage[pretest,nocommandblocks]{progreg}
\usepackage[bracketinterpret,prefixflatinterpret,bracketmodalinterpret,fixformat,differentialdL]{dL}

\let\citet\cite
\newif\iflongversion
\longversiontrue

\usepackage{booktabs} 
\providecommand{\hfilll}{\hskip 0pt plus 1filll}
\def\linferenceRuleSidenoteSeparation{~~\hfilll}
\renewcommand*{\irrulename}[1]{\text{\textcolor{gray}{#1}}}

\usepackage{prettyref}
\newcommand{\rref}[2][]{\prettyref{#2}}
\newrefformat{ch}{Chapter\,\ref{#1}}
\newrefformat{pt}{Part\,\ref{#1}}
\newrefformat{sec}{Section\,\ref{#1}}
\newrefformat{app}{Appendix\,\ref{#1}}
\newrefformat{def}{Def.\,\ref{#1}}
\newrefformat{thm}{Theorem\,\ref{#1}}
\newrefformat{prop}{Proposition\,\ref{#1}}
\newrefformat{lem}{Lemma\,\ref{#1}}
\newrefformat{rem}{Remark\,\ref{#1}}
\newrefformat{cor}{Corollary\,\ref{#1}}
\newrefformat{ex}{Example\,\ref{#1}}
\newrefformat{tab}{Table\,\ref{#1}}
\newrefformat{fig}{Fig.\,\ref{#1}}
\newrefformat{case}{case\,\ref{#1}}
\newrefformat{prn}{Principle\,\ref{#1}}
\usepackage{listings}
\newcommand{\VeriPhy}{VeriPhy\xspace}

\newcommand{\CdGL}{\textsf{CdGL}\xspace}

\newcommand{\bebecomes}{\mathrel{::=}}
\newcommand{\alternative}{\,|\,}
\newcommand{\ivr}{Q}
\newcommand{\lequiv}{\leftrightarrow}

\definecolor{eclipseBlue}{RGB}{42,0.0,255}
\definecolor{eclipseGreen}{RGB}{63,127,95}
\definecolor{eclipsePurple}{RGB}{127,0,85}

\lstdefinelanguage{kaisar}
{
  morekeywords={
    proves,    
    note,    let,
    ?,  !,
    /++,    ++/,    /--,    --/, ++,
    @,
    using,    by,        ...,
    init:,    old:,    final:,    mid:,    body:,    ode:, one:, two:, theEnd:, adm:, monitor:,
    auto,    prop,    rcf, exhaustion,    hypothesis,    induction,    solution,    proof,    end,    guard,
    switch,    case,
    andI,    orE,
    for,
    print, pragma, option
  },
	  literate= {@}{{{\color{eclipsePurple}@}}}1 {?}{{{\color{eclipsePurple}?}}}1 {!}{{{\color{eclipsePurple}!}}}1 {!=}{{{\color{black}!=}}}2 {ode(t):}{{{\color{eclipsePurple}ode(t):}}}7 {ODEMAGICSTR}{{{\color{eclipsePurple}ode(t, acc):}}}8,
  alsoletter={@,?,!,+,-,*,/},  
  alsodigit={:},
  sensitive=true, 
  morecomment=[s]{/*}{*/}, 
  morestring=[b]" 
}

\lstdefinelanguage{bellerophon}
{
  morekeywords={
   Real,B,R,HP,
   Bool,
   let,
   def,tactic,
   @invariant,
   Lemma, Tactic, 
   Theorem,
   ArchiveEntry, ProgramVariables,
  Problem,
   SharedDefinitions,Definitions,
   End
  },
  sensitive=true, 
  alsoletter={@},
  morecomment=[l]{//}, 
  morecomment=[s]{/*}{*/}, 
  morestring=[b]" 
}

\begin{document}

\title{Structured Proofs for Adversarial Cyber-Physical Systems}

\author{Brandon Bohrer}
\orcid{0000-0001-5201-9895}
\affiliation{
  \department{Computer Science Department}
  \institution{Carnegie Mellon University}
  \city{Pittsburgh}
  \state{Pennsylvania}
  \postcode{15213}
  \country{USA}
}
\email{bbohrer@wpi.edu}
\authornote{This author is now affiliated with Worcester Polytechnic Institute.}

\author{Andr\'e Platzer}
\orcid{0000-0001-7238-5710}
\affiliation{
  \department{Computer Science Department}
  \institution{Carnegie Mellon University}
  \city{Pittsburgh}
  \state{Pennsylvania}
  \postcode{15213}
  \country{USA}
}
\email{aplatzer@cs.cmu.edu}
\thanks{This article appears as part of the ESWEEK-TECS special issue and was presented in the International Conference on Embedded Software (EMSOFT), 2021.}
\begin{abstract}
Many cyber-physical systems (CPS) are safety-critical, so it is important to formally verify them, e.g. in formal logics that show a model's correctness specification \emph{always} holds. \emph{Constructive Differential Game Logic} (\CdGL) is such a logic for (constructive) hybrid games, including hybrid systems. 
To overcome undecidability, the user first writes a proof, for which we present a proof-checking tool.

We introduce \emph{Kaisar}, the first language and tool for \CdGL proofs, which until now could only be written by hand with a low-level proof calculus.
Kaisar's \emph{structured proofs} simplify challenging CPS proof tasks,
especially by using programming language principles and high-level stateful reasoning. 
Kaisar exploits \CdGL's constructivity and refinement relations to build proofs around models of game strategies. The evaluation reproduces and extends existing case studies on 1D and 2D driving. Proof metrics are compared and reported experiences are discussed for the original studies and their reproductions.
\end{abstract}


\maketitle

\section{Introduction}
\label{sec:kaisar-introduction}
Cyber-physical systems (CPSs), where embedded computers control physical devices, are often safety-critical. 
Important examples of safety-critical applications include robotics, automotives, aviation, spaceflight, medical devices, and power systems.
Formal methods for CPS are essential to ensuring crucial correctness properties of system models and the implementations of CPS on embedded processors.
Among formal methods, theorem-proving approaches are essential both because they provide the high degree of rigor that CPSs demand and because they can show correctness of a model for \emph{all} of its uncountably many behaviors and states, in stark contrast to testing-based methods, which inherently test only finitely many behaviors and states. This exhaustiveness allows catching bugs that are difficult to find with other approaches and allows catching them early, in the design stages of the development workflow, when fixing bugs is cheap. It also allows proving their absence once corrected, upon which the proof serves as strong evidence that illustrates a design's correctness. When combined with synthesis, theorem-proving continues to ensure correctness throughout the implementation stages of a development workflow~\cite{DBLP:conf/pldi/BohrerTMMP18}.

Theorem-proving in the \emph{differential dynamic logic} (\dL)~\cite{DBLP:journals/jar/Platzer08,DBLP:journals/jar/Platzer17,Platzer18} family of logics is of particular note, because it has been successfully applied to a variety of CPS case studies~\cite{Platzer18} modeled as hybrid systems, including case studies in automotives, aviation, and medical robotics. The \KeYmaeraX~\cite{DBLP:conf/cade/FultonMQVP15} prover implements \dL and its generalization to hybrid games, \emph{differential game logic} (\dGL)~\cite[Ch.\ 14]{Platzer18}\cite{DBLP:journals/tocl/Platzer15}. The logics \dL and \dGL are notable both because they can analyze polynomial ordinary differential equations (ODEs) that have no closed-form solution~\cite[Ch.\ 10-11]{Platzer18}\cite{DBLP:journals/jacm/PlatzerT20} and because they can prove a broad class of safety and  liveness properties that ensure functional correctness of a model. For example, safety properties of a transportation CPS include collision-freedom, while liveness properties include reaching a destination in finite time. Games further excel at modeling CPSs whose environments are adversarial, as well as proving \emph{reach-avoid} properties: liveness goals are eventually reached while staying safe throughout. By proving correctness of a system which operates in an adversarial environment, game proofs also amount to proofs 
of security against an adversary who attacks the system to try to violate correctness. This paper considers \emph{Constructive Differential Game Logic} (\CdGL)~\cite{DBLP:conf/cade/BohrerP20}, the recent \emph{constructive logic} counterpart of \dGL. Its constructive foundations notably
simplify synthesis (or extraction) of correct monitoring and control code~\cite[Ch.\ 8]{Bohrer21} by giving them an immediate theoretical basis. 
We assume no familiarity with constructive logic, 
but the constructive notion of proofs as programs is exploited in Kaisar's design.
Broadly, synthesis is a key motivation, but its details are discussed elsewhere~\cite[Ch.\ 8]{Bohrer21} to save space.

Hybrid games are applicable to a broad range of applications and adversaries.
This paper demonstrates the usefulness of hybrid games for the correctness and security of adversarial CPSs via case studies in autonomous driving and ground robotics, which are important examples of embedded systems.
Our examples include adversarial timing and actuation, where the adversary has the (bounded) power to disturb both a scheduler and a vehicle's motion.
Related literature on CPS verification~\cite{Platzer18} indicates that the modeling and proof techniques we provide generalize across the aforementioned domains. 
Likewise, our approach generalizes to adversaries which manipulate \emph{any} aspect of hybrid games, e.g., timing, sensing, control, and physics.

The undecidable complexity of hybrid systems results in different tradeoffs for every formal method.
The tradeoff for expressive logics like the \dL family is that human proof insight is required for proofs to scale. In contrast, reachability methods typically achieve greater automation by limiting the class of supported models and properties, or by approximating the system dynamics.

A user's proof insights are expressed in a proof language and checked for correctness by a tool. Proof language design is a significant research topic in its own right. Due to the subtleties of CPS, proof-writing and revisions can take longer than writing an initial (faulty) controller and model, motivating the design of languages which increase productivity of verification, especially productivity of revisions and maintenance of the CPS and its correctness proof.

We introduce a new CPS proof language, so we necessarily discuss logic and language design at length. 
Yet, the unique strengths of proofs versus other approaches for achieving CPS correctness make proof languages relevant to the broader CPS and embedded systems community. 
All who develop CPS and embedded systems have a vested interest in the systems' correctness, 
thus novel verification technology is relevant to all.
Though formal proofs can present a learning curve for the broader community, this fact 
reiterates the value of language-based verification approaches, because each advancement in language design might help flatten verification's learning curve.

This paper presents a standalone structured proof language and tool named \emph{Kaisar},
the first proof language for \CdGL.
At the highest level, Kaisar's novel design follows from two core design principles which respectively come from \CdGL's constructivity and structured proof languages (\rref{sec:kaisar-relwork}):
\begin{inparaenum}[(I)]
\item \label{prn:curry-howard}
The \emph{Curry-Howard} isomorphism for games says that a \emph{constructive} game proof consists of a programmatic strategy, which behaves like a hybrid system, and a correctness proof for that strategy.
\item \label{prn:stability}
Proofs should be \emph{stable} in the sense that a revision to one model or proof statement will not require non-local changes to conceptually unrelated statements.
\end{inparaenum}

These principles manifest in features which support design priorities including readability, maintainability, ease of expression, and flattening the learning curve. Given the current state of the art (e.g., the language Bellerophon~\cite{DBLP:conf/itp/FultonMBP17} for unstructured \dL proofs in the \KeYmaeraX~\cite{DBLP:conf/cade/FultonMQVP15} theorem prover), these priorities remain crucial to productivity and accessibility of new languages.
Since it is difficult for experiments to conclusively assess the above design priorities, our empirical evaluation (\rref{sec:kaisar-evaluation}) uses incomplete metrics as proxies, such as counts of total or changed proof lines.
Those metrics are crucially supplemented with ample examples by which the reader may assess readability and with discussion of how our novel design decisions pursue the goals.

\rref{prn:curry-howard} is the organizing principle for our workflow's learning curve.
Organizing proofs around strategy programs with inline annotations both allows easy visual tracing of the relationship between proof and code and also lets us exploit  insights on readability, maintainability, and scalability from programming language design. 
First, the user simply writes a hybrid game strategy, analogous to a hybrid system.
Next, specifications are inserted gracefully and iteratively into the model, then Kaisar attempts automatic proofs.  If models are simple, the hardest task, manual proof, can be completed avoided. More often, manual proofs are used, but only for crucial questions and only once easier tasks are completed.  Lastly, the approach is extended from strategies to games with automatic \emph{refinement reasoning}~\cite{DBLP:conf/rta/BohrerP20} that checks whether a given strategy plays a given game. This ability to build game verification on top of systems verification is crucial (because games are avant-garde) and also key to Kaisar's learning philosophy. 
We do not pursue broader accessibility by removing the need for proof expertise, but by designing Kaisar to simplify acquiring that expertise.

\rref{prn:stability} is intentionally broad because stateful systems like hybrid games require notions of proof stability which subsume, yet exceed, those in common use.
Structured proofs (\rref{sec:kaisar-relwork}) support stability with features like giving persistent names to proven facts for future reference and defining reusable helper functions that decompose large models. 
Kaisar additionally introduces a new feature, \emph{labeled reasoning}, which provides stable references to past and future system \emph{states}.

Labeled reasoning significantly simplifies a variety of particularly important CPS paradigms:
\begin{inparaenum}[\it i)]
\item model-predictive controllers~\cite{DBLP:journals/ijrr/MitschGVP17} that predict future states to make safe control choices,
\item ODE invariant~\cite{DBLP:journals/ijrr/MitschGVP17} proofs that compare current and initial states,
\item stability proofs~\cite{chan2016formal} that track change in distance,
\item liveness~\cite{DBLP:journals/ijrr/MitschGVP17} and reach-avoid \cite{DBLP:conf/cade/BohrerP20} 
 proofs which require progress arguments,
\item sandbox controllers~\cite{DBLP:conf/pldi/BohrerTMMP18} that allow the model-predictive approach to safely interact with external control code,
which is key for synthesis~\cite{Bohrer21}.
\end{inparaenum}

We prioritize presenting the Kaisar language and its uses over presenting implementation details,
but we briefly note that the implementation of Kaisar's novel design overcomes novel technical hurdles.
Because hybrid games have mutable state to accurately reflect the state changes of dynamical systems, \rref{prn:stability} and especially labeled reasoning require a systematic treatment of past and future state.
The extended version of this work~\cite[Ch.\ 7]{Bohrer21}, describes that treatment:
in combination with rich data structures for organizing assumptions and definitions, a
notion of  \emph{static single-assignment} proof is used which is inspired by the standard notion from compilers~\cite{DBLP:journals/toplas/CytronFRWZ91}, but serves a distinct purpose of systematically naming and remembering historical program state for easy processing by high-level proof automation.

Related work, including comparisons of Kaisar against other structured proof languages and against Bellerophon, is in  \rref{sec:kaisar-relwork}.
The connection between games and systems is explained in \rref{sec:kaisar-relationships}.
The heart of the paper, \rref{sec:kaisar-by-example}, introduces Kaisar. It starts with toy examples but builds to fundamental paradigms.   Kaisar is evaluated against Bellerophon, \KeYmaeraX's unstructured language for proofs, on existing case studies in \rref{sec:kaisar-evaluation}. 
Future work is  in \rref{sec:kaisar-future-work}.

\section{Related Work}
\label{sec:kaisar-relwork}
We first discuss formal methods for CPS besides theorem proving, then languages related to Kaisar. 
For space reasons, we only cite a representative subset of related works.

\paragraph{Formal Methods for CPS}
Hybrid systems, a fragment of hybrid games, are  canonical models of  CPS because they can model both discrete and continuous change.  The two main categories of (offline) verification methods for hybrid systems are reachability analysis and theorem-proving.
Reachability analysis typically achieves a higher degree of automation, but restricts the class of allowed models or correctness properties in order
to provide scalability. To give a few prominent examples, SpaceEx~\cite{DBLP:conf/cav/FrehseGDCRLRGDM11} scales to ODEs with over 100 variables, but does so by assuming ODEs are affine (${\approx}$linear) and using conservative overapproximations of (sets of) system states.
Unbounded-time safety guarantees are only supported when they can be shown over the conservative dynamics.
Flow*~\cite{DBLP:conf/cav/ChenAS13} is notable for supporting non-linear ODEs, but only supports bounded-time safety for conservative state representations. Theorem-proving approaches such as \dL~\cite{Platzer18} and \CdGL~\cite{DBLP:conf/cade/BohrerP20} can show unbounded-time safety and liveness of non-linear ODEs with exact state representations at the cost of less automation, thus greater reliance on human proof insights to enable scalability.

Other major formal methods topics include \emph{runtime verification}~\cite{Krogh1998TheSA} and \emph{synthesis}.
Runtime verification shows correctness via runtime checks. 
Synthesis is the  generation of correct code from a model, specification, and sometimes a proof.
For CPS, runtime verification and synthesis have been combined. For example, \VeriPhy~\cite{DBLP:conf/pldi/BohrerTMMP18} generates controllers that \emph{sandbox} an untrusted controller by replacing any potentially-unsafe action with proven-safe ones. \VeriPhy yields a highly formal, machine-checked proof that an implementation is correct. 
The first author's thesis~[Ch.\ 8]\cite{Bohrer21} builds on the present paper to provide a novel synthesis tool for Kaisar in a similar approach.

Code is rarely synthesized \emph{from} games, and even then only small fragments~\cite{DBLP:conf/concur/HenzingerHM99} are addressed.
More often, games are used internally to implement synthesis for hybrid systems~\cite{tomlin2000game}.
As with theorem proving and reachability, hybrid systems synthesis approaches make tradeoffs between expressiveness, decidability, and scalability, e.g., \VeriPhy-style synthesis requires a proof to overcome decidability, but supports non-linear ODEs and provides rigorous machine-checked proofs of correctness for generated code.
Synthesis approaches without reliance on human proofs~\cite{DBLP:conf/emsoft/TalyT10,tomlin2000game,DBLP:conf/concur/HenzingerHM99} typically limit the class of hybrid systems or accept that synthesis will not always succeed.

We now discuss three classes of languages related to Kaisar.
The classes are not mutually exclusive.
Kaisar combines all three and other provers have combined structured and unstructured proofs~\cite[\S2.2.2]{DBLP:journals/jfrea/GrabowskiKN10} or combined unstructured proofs with
limited annotation support~\cite{DBLP:conf/cade/FultonMQVP15}.

\paragraph{Structured Proofs.}
Structured proofs allow a high degree of generality and control  without sacrificing readability.
Mizar~\cite{DBLP:journals/jfrea/GrabowskiKN10} is the canonical structured proof language and Isar~\cite{Wenzel07isabelle/isar} is another prominent example. 
Structured features can include fact naming, function definitions, declarative block structure, and
explicit syntax for low-level proof steps. 

Kaisar innovates relative to other structured languages in its foundations, design, and implementation.
Whereas Mizar and Isar are respectively founded in set theory and higher-order logic,
Kaisar is founded in a program logic, 
whose rich, frequent, and non-trivial changes in mutable state demand a novel design and implementation.
Questions of state, though of broad interest, are certainly of special interest to CPS, which fundamentally rely on state changes resulting from their underlying discrete and continuous dynamics.
A key novelty of Kaisar's is its concise and maintainable notation for reasoning across states.
Kaisar's implementation is equally novel because it automatically manages state to ensure soundness.
For example, a true property may later become false if some variable it mentions is reassigned, so Kaisar automatically distinguishes past and current truth.
Such careful state managament is essential \emph{any} time proof and mutation are intermixed, but it 
culminates in labeled reasoning, Kaisar's novel structuring principle for stateful systems.
Labeling shows its novel impact by simplifying proofs of common CPS proof idioms (\rref{sec:kaisar-idioms}).

Compared to previous structured languages, a second major innovation of Kaisar is its built-in support for games and specifically its built-in automation for refinement proofs which reduce hybrid game verification to hybrid system verification. Though the logics underlying Mizar and Isar are generic enough to \emph{formalize} games~\cite{DBLP:conf/cade/Platzer19} and refinements~\cite{DBLP:journals/afp/ArmstrongGS14}, neither system provides any special automation for games or refinements. Such automation is crucial to Kaisar's approach toward a flat learning curve because it reduces the learning of game verification to two simpler tasks: learning systems verification and learning (automated) refinements.

\paragraph{Annotation-Based Verification.}
Annotation-based languages allow gradually adding concise, high-level specifications to existing models or programs, but can only scale when verification of those specifications is supported by additional features. Annotation-based proofs are founded in \emph{proof outlines}, introduced by Owicki~\cite{DBLP:books/garland/Owicki75} and further studied by Apt et.\ al.~\cite{apt2010verification}.
Proof outlines annotate programs with assumptions, invariants, and optionally with assertions that must hold at a given intermediate point.
Canonical examples of tools using this approach include the ESC family~\cite{DBLP:conf/procomet/Leino98}. 

\KeYmaeraX provides limited annotations for CPS proofs: loops and ODEs can be annotated with invariants 
that are consumed by a fully-automatic proof procedure. The crucial limitation is that the verification paradigm changes entirely once fully-automated proofs fail: the user writes a proof in Bellerophon~\cite{DBLP:conf/itp/FultonMBP17} or equivalently interacts with a user interface that generates the proof script.

\paragraph{Tactics and Unstructured Proofs.}
Other relevant languages include tactic languages~\cite{Delahaye:2000:TLS:1765236.1765246,malecha2015rtac-coqpl,Ziliani:2013:MMT:2544174.2500579} (often capable of implementing automation) and unstructured proof languages~\cite[Ch.\ 7]{DBLP:phd/dnb/Grebing19}\cite{DBLP:conf/itp/FultonMBP17} (for writing concrete, individual proofs).
The prover's implementation language may be used, or a domain-specific language may be provided.
We give special attention to \KeYmaeraX's Bellerophon~\cite{DBLP:conf/itp/FultonMBP17} proof language, because its underlying logics \dL and \dGL are direct ancestors of the logic \CdGL which Kaisar targets and because Bellerophon is applied to the same application domain: CPS.
Though the application domains and underlying logics of Bellerophon and Kaisar are intimately related, the two languages could not be more different, when considered as languages.

The core Bellerophon language is barebones and combines built-in proof procedures using regular-expression style operations.
For example, sequencing allows one proof step to be applied after another, repetition allows applying the same method multiple times, and alternation allows several different proof attempts to be tried, until one succeeds.
In contrast, Kaisar has a richer core language, whose key concepts include freely mixing models with proofs, freely mixing automatic and manual proof styles, freely introducing names and definitions, and the new labeling feature.

By building powerful language constructs into its core design, Kaisar makes its high-level design goals far easier to achieve, 
in support of key practical goals such as ease of learning, readability, and maintainability. Thus the design differences between Bellerophon and Kaisar are not merely of narrow linguistic interest, but relate directly to Kaisar's core practical goals. We enumerate several important practical benefits of Kaisar's language-driven approach.

Firstly, Kaisar provides \emph{traceability}, i.e., Kaisar makes the relationship between proof statements and model statements immediately obvious: 
because proofs are written inline with model statements, it is always clear \emph{which} model statement is proved by a given proof statement. Traceability is a crucial aspect of readability, yet it is not provided in Bellerophon, where proofs and models are completely separate from one another and can differ significantly in their structure and content.

Secondly, by choosing core language concepts which are broadly agreed to promote maintainability, Kaisar greatly increases the odds that maintainable models and proofs are written in practice and not only in theory. For example, fact naming is a core Kaisar language feature that Bellerophon does not provide.
The core Bellerophon design refers to facts using numeric indices, which are highly unstable during proof maintenance (e.g., in case studies mentioned in \rref{sec:kaisar-evaluation}) and do not provide useful high-level information to a reader, and are thus far less likely to result in maintainable proofs when compared to fact names that are stable during maintenance and can communicate high-level information.
Because the core Bellerophon language design refers to facts unmaintainably, alternative features must be added after-the-fact.
For example, the latest release of Bellerophon builds optional search-based references on top of numeric ones: the proof author can ask Bellerophon to search for the index of a suitable formula instead of writing the index manually.
When high-level features are added after-the-fact, they remain optional. 
Because older features are typically better-known and better-documented, they are likely to see continued use from users who stand to benefit from switching to new features.
In short, Kaisar makes entire classes of bad proofs impossible, whereas Bellerophon merely makes good proofs possible for expert users.

Thirdly, Kaisar's language-based approach enables advanced features which would be far more difficult to implement without special language support.
Kaisar provides to the language implementation a global view of the model and proof, including detailed information on how program state changes over time. This global view is used to give facts persistent names which can be soundly accessed even after the program state changes and is also used to implement the new labeled reasoning feature. Both features would be difficult to implement without this global view, which is why neither feature was provided in Bellerophon.
The advantages of the language-based approach are particularly strong when several advanced features must co-exist: though Bellerophon and Kaisar both allow definitions, Kaisar's rich language data structures allow freely combining those definitions with labeling and naming, which would be hard to implement otherwise.

In contrast to the highly general notion of backward and forward labeled references provided by Kaisar, its predecessors provided only limited special cases of historical reference because they lack the rich data structures that make Kaisar's labeling feature possible.
There are many tools with limited historical reference features, of which we cite only a selection~\cite{ 
DBLP:conf/cade/FultonMQVP15,DBLP:books/garland/Owicki75,apt2010verification,DBLP:books/daglib/0067105} and discuss Bellerophon specifically, because the limitations of the different tools are similar.
In Bellerophon, the notation \texttt{old(e)} stands for the value of the expression \texttt{e} at the start of the program being proved in the current proof state.
References to arbitrary previous states are not supported directly, nor are references to hypothetical future states.
As we discuss in \rref{sec:kaisar-idioms}, hypothetical future references are crucial to expressing a variety of reusable CPS proof idioms in Kaisar.
Because Kaisar's labeling feature enables important classes of proof approaches which past tools did not support, it is an entirely new feature in the qualitative sense, i.e.,
it is so much more general that it can be (and is) used in notable ways that its predecessors cannot.

Though the contributions of Kaisar compared to Bellerophon are significant, we note that Bellerophon also contributed important features when compared to general-purpose proof languages and that those past contributions have helped to make Kaisar possible.
Compared to general-purpose provers, Bellerophon is notable because it provides a broad library of domain-specific features for proofs about CPS.
Those features range from low-level sequent calculus proofs to automated proof search and invariant reasoning.
Though Kaisar also broadly supports low-level manual proofs, automated proof search, and invariant-style proof, Bellerophon enjoys the benefits of age and thus currently provides a more mature standard library. 
For example, Bellerophon has access to advanced invariant search methods~\cite{DBLP:journals/fmsd/SogokonMTCP} which are not yet provided in Kaisar.

In summary, Kaisar's language design provides major novel features when compared to any of its predecessors, including previous structured languages as well as Bellerophon.
These novel features are not incidental, but rather they directly support Kaisar's goals of providing a gentler learning curve for CPS verification and supporting readability and maintainability.

\section{Connecting Games to Systems}
\label{sec:kaisar-relationships}
We connect Kaisar and \CdGL~\cite{DBLP:conf/cade/BohrerP20}.
Kaisar is \CdGL's first proof language and tool. Previously, \CdGL only had a low-level proof calculus for paper proofs.
Knowledge of \CdGL foundations is not assumed in later sections, but this section is aimed at readers interested in the foundations.
Formulas $\phi, \psi, \rho$ of \CdGL describe existence of computable, non-random winning strategies for 2-player, zero-sum, perfect-information hybrid games $\alpha.$
Such winning conditions are used to express correctness theorems.
\CdGL formulas also include statements about real-valued arithmetic terms $f,$ which are often a key part of correctness statements.
Terms $f,$ which also appear in games, allow polynomials, division, and user-defined functions.
We define the language of hybrid games as a recursive grammar.
The notation below says  games $\alpha, \beta$ are produced ($\bebecomes$) by freely combining each game construct. Vertical bars $(\alternative)$ are written between (the syntax of) each construct.
\[\alpha, \beta \bebecomes \humod{x}{f} \alternative \prandom{x} \alternative \pevolvein{\D{x}= f}{\ivr} \alternative \ptest{\phi} \alternative \alpha;\beta \alternative \alpha\cup\beta \alternative \prepeat{\alpha} \alternative \pdual{\alpha}\]
Game $\humod{x}{f}$ deterministically assigns variable $x:\reals$ the value of term $f$.
Game $\prandom{x}$ lets the player pick the value (in $\reals$) of $x$ nondeterministically.
Game $\pevolvein{\D{x}=f}{\ivr}$ evolves the ODE $\D{x}=f$ for a time $0 \leq d \in  \reals$ chosen by the player, where the player proves $\ivr$ holds continuously until time $d$.
Game $\ptest{\phi}$ is a no-op if the player can prove $\phi,$ else they lose immediately.
Game $\alpha;\beta$ is $\alpha$ followed by $\beta$.
In $\alpha\cup\beta,$ the player picks which of $\alpha$ or $\beta$ to play.
In game $\prepeat{\alpha}$, the player chooses after each play of $\alpha$ whether to play another round.
Game $\pdual{\alpha}$ plays $\alpha$ with the two players reversed.

The players' standard names are Angel and Demon.
In \CdGL, Angel is \emph{our} player, controlled computably, while Demon is our adversary.
The key formulas of \CdGL are $\ddiamond{\alpha}{\phi}$ and $\dbox{\alpha}{\phi},$
 which both mean we (Angel) win $\alpha$ with postcondition $\phi$ by playing a constructive strategy.
 They differ in that we (Angel) control all top-level choices in $\ddiamond{\alpha}{\phi}$ while the opponent (Demon) controls all top-level choices in $\dbox{\alpha}{\phi}.$ Colloquially, Angel ``moves first'' in the former case and Demon ``moves first'' in the latter. Dualities switch control between players, e.g., Demon resolves the value of $x$ in $\ddiamond{\pdual{\{\prandom{x}\}}}{x^2 \geq 0}$.
For a given logic, a proof system is sound  if all proofs it accepts have \emph{valid} conclusions. 
The notion of \emph{validity} differs between logics.
We define Kaisar's soundness relatively: all proofs it accepts must have conclusions provable in the (sound) \CdGL~\cite{DBLP:conf/cade/BohrerP20} proof calculus.

Following the \emph{Curry-Howard isomorphism for games}~\cite{DBLP:conf/cade/BohrerP20},
Kaisar proofs consist of \emph{executable strategies} for \CdGL games and proof annotations.
Kaisar strategies differ from game syntax by replacing the duality operator with new high-level Angelic constructs.
Kaisar strategies are no less general than \CdGL games, but first we show how \CdGL theorems are inferred from Kaisar proofs.

To find the \CdGL formula proved by a Kaisar strategy, Kaisar automatically reads off a game $\alpha,$ called the strategy's \emph{game reification}, and a postcondition $\phi,$  such that the strategy proves the \CdGL formula  $\dbox{\alpha}{\phi}$.
The crucial step is finding $\alpha.$
We define $\alpha$ as the game that forces Angel to follow her given strategy,
but makes no new restrictions on Demon's strategy since he is an adversary.
The restriction to shape $\dbox{\alpha}{\phi}$ does not lose generality due to the equivalence $\dbox{\pdual{\alpha}}{\phi} \lequiv \ddiamond{\alpha}{\phi}$.

In the most basic Kaisar workflow, the user never needs to write a theorem statement in \CdGL.
After writing their strategy, the user can write the command \texttt{conclusion proofName;} which will automatically compute \texttt{proofName}'s game reification and display the proven \CdGL theorem. This automation is important for reducing Kaisar's learning curve because it allows a user to start using the Kaisar proof language with no prior knowledge of the \CdGL logic.

By inferring \CdGL theorems from Kaisar proofs, we have connected Kaisar to \CdGL, but
the converse connection also matters: can every (low-level) \CdGL proof be written in Kaisar, so that Kaisar is fully general?
The practical answer is ``yes,'' except for a few advanced proof rules~\cite[Rule\ DV]{DBLP:conf/cade/BohrerP20}.
We discuss Kaisar's generality via discussion of Kaisar's \texttt{proves} command. 
Once  \texttt{theProof} has been checked with its default \texttt{conclusion},
writing \texttt{proves theProof "[}$\alpha$\texttt{]}$\phi$\texttt{";}  asks Kaisar whether  $\dbox{\alpha}{\phi}$ is
(another) valid conclusion of \texttt{theProof}. To reduce the learning curve, \texttt{proves} is optional, meant for users who find the generated \texttt{conclusion} too  verbose.

The \texttt{proves} command is built on automated refinement reasoning~\cite{DBLP:conf/rta/BohrerP20}, which allows
generalizing theorems about one game to a different game.
The key rule is refinement elimination~\cite[Rule\ R\m{[\cdot]}]{DBLP:conf/rta/BohrerP20}, which concludes $\dbox{\beta}{\phi}$ from the known fact $\dbox{\alpha}{\phi}$ for any game $\beta$ which $\alpha$ refines. To apply the rule and prove $\dbox{\beta}{\phi},$ the crucial step is to check whether $\alpha$ refines $\beta$.

Automated refinement-checking follows game and proof structure, 
letting them differ by simple game-algebraic~\cite{DBLP:journals/sLogica/Goranko03} laws like distributivity (\cite[Rule\ \m{;d_R}]{DBLP:conf/rta/BohrerP20}).
Strategies' leaves may use more precise connectives than the game, but not vice-versa. 
Notably, deterministic assignments $\humod{x}{f}$ are reused as strategies of Angelic assignments $\pdual{\{\prandom{x}\}}$. 
Differences in assignments are one way that a game can admit many different winning strategies. 
When the strategy contains statements (often assertions of lemmas) not appearing in the game, Kaisar's \emph{ghost} statements (\rref{sec:kaisar-ghost-proof}), which serve as justifications but are not interpreted as code, let the refinement checker soundly erase statements.
These refinement methods suffice in both practice and theory; 
we explain practice first.

The above methods suffice in practice because 
a typical \CdGL proof roughly follows the shape of the game it proves.
If \verb|proves| fails, the failure's reason and location are reported. Most failures are fixed by adding ghosts or ensuring the model and proof make the same assumptions. If refinement still fails, it may indicate that the user  truly wrote a strategy that plays the wrong game.

It would suffice as theoretical justification of Kaisar's generality to show that for every proof of a \CdGL formula $\dbox{\beta}{\phi},$ there exists a Kaisar strategy whose game reification $\alpha$ can be proven to satisfy postcondition $\phi$ and refine $\beta,$ using the same refinement rules provided by our refinement checker. The theory of refinement~\cite{DBLP:conf/rta/BohrerP20} goes further by producing a notion of \emph{system reification}, i.e., it produces game reifications which happen to be proper hybrid systems (1-player games).

It is a theorem~\cite[Thm.\ 10]{DBLP:conf/rta/BohrerP20} that for \emph{every} proof of a \CdGL modality $\dbox{\beta}{\phi},$ game $\beta$ is refined by the system reification $\alpha$.
It is also a theorem~\cite[Thm.\ 9]{DBLP:conf/rta/BohrerP20} that $\dbox{\alpha}{\phi}$ is provable in \CdGL, i.e., the game postcondition also holds of the system reification.
Crucially, the proof of each consists of an algorithm that explicitly demonstrates which rules are needed to check refinements and which \CdGL rules are used to prove $\dbox{\alpha}{\phi}$.
The refinement-checking algorithm provided by the proof is the direct inspiration for Kaisar's refinement checker and the \CdGL rules used are, with minor exceptions~\cite[Rule\ DV]{DBLP:conf/cade/BohrerP20}, also provided by Kaisar.
In summary, every \CdGL paper proof can be divided into a proof about systems, which are a subset of strategies, and a refinement proof. In typical use, each proof uses techniques available in Kaisar, thus \CdGL proofs can in principle be mechanically translated into Kaisar strategies and the use of Kaisar's refinement checker.

Though this argument for Kaisar's generality was nontrivial, Kaisar's refinement-based paradigm itself has the key strength of smoothing Kaisar's learning curve. Strategies, which act like hybrid systems, can be verified first, then results can be generalized to hybrid games after the fact.

Even though game proofs can be reified as systems, games provide unique advantages in Kaisar compared to systems. Games support stronger separation between model and proof; for example, controller models can concisely list the available actions but defer the determination of each action's safety conditions to the proof. As control complexity grows, this separation makes models shorter, thus typically easier to read, communicate, and trust. 
On the flip side, verification 
is more mature for hybrid systems than for games, 
so Kaisar's access to hybrid systems techniques is a benefit.

The \CdGL refinement calculus is at least as expressive as \CdGL, thus undecidable~\cite{DBLP:conf/rta/BohrerP20}.
Because the \verb|proves| command should always terminate, it uses an incomplete, yet flexible refinement checker.

\section{Kaisar by Example: Proving Hybrid Games}
\label{sec:kaisar-by-example}
We present the Kaisar language in detail by giving examples which start with toy proofs and conclude with demonstrations of major idioms: (logical) model-predictive control, sandboxing, and reach-avoid correctness.
As we discuss the examples, we discuss crucial issues of logical \emph{soundness}, i.e., ensuring that only correct proofs are accepted.
For full-scale case studies, see \rref{sec:kaisar-evaluation}.
 
We use the adjective \emph{Angelic} for things we control and \emph{Demonic} for things we do not, following the names Angel for our player and Demon for the opponent, respectively.
Following the Curry-Howard isomorphism for games~\cite{DBLP:conf/cade/BohrerP20},
Kaisar proofs prove that Angel computably \emph{wins} some hybrid game, meaning she reaches the game without failing any tests, assuming Demon passed his tests as well.
If Demon fails a test, Angel wins immediately.
The word \emph{fact} refers to a formula that is true, either by proof or because it was assumed.

\subsection{Core Propositional Connectives}
\label{sec:kaisar-prop-proofs}
We begin with propositional operators.
Each \verb|?| statement, such as in the first line below, means an assumption.
Assumption statements are \emph{Demonic tests} in that Demon loses if the test condition is not provable. In this syntax, \verb+x = 0+ and  \verb+x = 1+ are formulas and \verb|bit| is a fact name. Dually, an \emph{Angelic test} is an assertion, written with the \verb|!| symbol. Angel loses an Angelic test if she cannot provide a constructive proof for it. Kaisar will attempt an automatic proof first, but the user must  manually justify the assertion with an unstructured proof (\rref{sec:kaisar-unstructured-step}) if automation does not suffice.

Symbol \verb|++| is plaintext notation for Demonic choice ($\cup$): when the game is played, our opponent chooses which of the two branches are played.
Kaisar's treatment of named facts handles choices automatically, but soundly: because \verb|bit| is a different formula in each branch.
When \verb|bit| is mentioned after the end of the choice, it refers to the (constructive) disjunction \verb+x=0 | x=1+ because Angel did not get to choose which branch was taken by Demon. We call such lookups \emph{disjunctive lookups} because they disjoin facts across branches.
If $x$ is neither 0 nor 1, then Demon fails a test regardless which branch he takes, at which point the game terminates and Angel wins.
Next is a \verb|switch| which accepts as its argument the disjunction \verb|bit|.
A \verb|switch| may omit its argument, in which case Kaisar attempts an automated proof that the disjunction of branch guards is constructively valid. This example needed an argument because the guards in the subsequent cases do not cover all program states, let alone constructively.
A \verb|switch| is an Angelic choice strategy: once it is proved that some branch's guard holds, Angel's strategy is to choose to play the \emph{first} such branch. Each \verb|case| has a guard formula, which acts as an assumption within the branch. Guard formulas optionally support the same fact naming notation \verb|name:(formula)| used in assumptions. The example below demonstrates Angelic and Demonic tests and choices by proving that  $0$ and $1$ are both nonnegative.

\begin{lstlisting}[language=kaisar]
{?bit:(x = 0); ++ ?bit:(x = 1);}
switch (bit) {
 case (x = 0) => !nonneg:(x >= 0);
 case (x = 1) => !nonneg:(x >= 0);
}
\end{lstlisting}

While the above \verb|switch| example uses a fact variable as the argument, more general arguments are permitted, called \emph{proof terms}, of which fact variables are one kind. The proof term language~\cite[\S7.3.1]{Bohrer21} has a steeper learning curve than other Kaisar constructs but provides a high degree of control by explicitly applying proof rules to their arguments. For that reason, the most crucial practical use of proof terms is when the user has a proof in mind, but automated proof has failed. Proof terms also appear in \verb|note| statements. We introduce \verb|note| next, along with the \verb|let| statement.

The statement \verb|note <name> = <proofTerm>| gives a fact \verb|<name>| to the conclusion of a \verb|<proof term>|.
Proof terms~\cite[\S7.3.1]{Bohrer21} use function-like syntax to apply proof rules to available facts.
The next example will use the \verb|andI| rule to conjoin two facts \verb|left| and \verb|right|.
A side benefit of \verb|note| statements is that their conclusions can be computed automatically and thus need not be written down.
Even for facts which could be proven with an assertion, \verb|note| statements are sometimes used if writing the conclusion would be tedious.

The \texttt{let} statement gives a name to term-level, formula-level, and game-level definitions, indicated by the symbols \verb|=|, \verb|<->|, and \verb|::=|, respectively.
The \texttt{let} statement is a core building block for code reuse, thus promoting readability and maintainability as well.
Here, \verb|square(z)| defines $z^2,$ duplicating the built-in exponentiation operator \verb|z^2| for example purposes.
If the right-hand side of a \verb|let| mentions variables other than the arguments, those variables take their values from the state where the definition is invoked.
The example \verb|note| below gives formula \verb|x < 0 & square(y) > 0|, which is the conclusion of the conjunction introduction rule \verb|andI|, the fact name \verb|both|.
Rules, including \verb|andI|, are described elsewhere~\cite[\S7.3.1]{Bohrer21}.

\begin{lstlisting}[language=kaisar]
let square(z) = z * z;
?left:(x < 0); ?right:(square(y) > 0);
note both = andI(left, right);
\end{lstlisting}
 
\subsection{Unstructured Proof Steps}
\label{sec:kaisar-unstructured-step}
In return for expressiveness, proofs often need human insight.
Assertions are no exception, because it is unknown whether the formulas of constructive arithmetic that typically appear in assertions are decidable~\cite{constructiveRealAlgebra}.
Unstructured proofs express user insights for assertions.
The simplest kind is \verb|by <method>|, for proof methods including:
\begin{align*}
\texttt{method} 
&\bebecomes  \texttt{auto} \alternative \texttt{prop} \alternative \texttt{rcf} 
\alternative \texttt{solution} \alternative \texttt{induction} \alternative \texttt{guard}
\end{align*}    
Method \texttt{auto} is the default when no unstructured proof is given; 
it combines the following \texttt{prop} and \texttt{rcf} methods.
Methods \texttt{prop} and \texttt{rcf} respectively use simple propositional rules or a first-order (\textbf{r}eal-\textbf{c}losed \textbf{f}ield) arithmetic solver. 
Methods \texttt{solution} and \texttt{induction} are only used in differential equation proofs (\rref{sec:kaisar-hybrid-games}).
The \texttt{guard} method is only used in \texttt{for} loops (\rref{sec:kaisar-for-loops}).

By default, a proof step uses all available assumptions which mention any of the conclusion's free variables.
Performance depends greatly on the number of assumptions, so it is crucial to allow choosing assumptions manually,
which we do by writing \verb|using <assumptions>| before \verb|by|.
To additionally use the default facts, write ellipses (\verb|...|) in the \verb|<assumptions>| list.
Assumptions can be proof terms as in \verb|note|, most often fact names.
As the below example shows, difficult arithmetic formulas sometimes have simple propositional proofs, in which case \verb|prop| is helpful.

\begin{lstlisting}[language=kaisar]
?a:(x = 0 -> y = 1); ?b:(x = 0 & ((z - x*w^2/(w^2+1))^42 >= 6));
!c:(y = 1) using a b by prop;
\end{lstlisting}

Note that the user is responsible for avoiding division by zero in \verb|x/y|.

\paragraph{Constructive Arithmetic}
Kaisar, like \KeYmaeraX~\cite{DBLP:conf/cade/FultonMQVP15},
supports automation of arithmetic proofs in Mathematica.
Classical arithmetic is decidable~\cite{Tarski}, but constructive decidability is unknown~\cite{constructiveRealAlgebra}, so Kaisar checks that constructive and classical truth agree before soundly applying Mathematica.

Specifically, we check for \emph{hereditary Harrop} formulas, a significant class where classical and constructive truth agree~\cite{DBLP:journals/apal/MillerNPS91}. A hereditary Harrop formula is one where the $\lor$ and $\lexists$ only appear in assumption positions.
\emph{Hereditary} means that, because implication is a ``negative'' connective, the notions of ``assumption'' and ``conclusion'' switch each time we look to the left of an implication, e.g., formula $(\phi \limply \psi) \limply \rho$ mentions $\psi$ in ``assumption position'' but neither $\phi$ nor $\rho$. Negations cause switching as well because they are defined as implications $\lnot \phi \lequiv (\phi \limply \bot)$. If the goal is not hereditary Harrop, \verb|prop|ositional automation and/or manual proof steps are used first, until any remaining goal is hereditary Harrop.

\subsection{Verifying Assignments}
\label{sec:kaisar-discrete}
We add assignments.
Deterministic assignment of term $f$ to variable $x$ is written $\humod{x}{f}$.
Demonic, nondeterministic assignments, where  Demon chooses the new value $x \in \reals,$ are written $\prandom{x}$.
As discussed in \rref{sec:kaisar-relationships}, strategies of \emph{Angelic} nondeterministic assignments are just deterministic assignments $\humod{x}{f}$.
Deterministic assignments can optionally use assumption-like syntax \verb|?id:(x := f);| where \verb|id| gets bound to an equality formula stating that the value of $x$ after the assignment equals the value of $f$ before.
Do not be misled by assumption-like assignment syntax. The assignment \texttt{?id:(x:=f);} updates the assigned variable $x$, thus it is distinct from an assumption statement \texttt{?id:(x=f);} that introduces an assumption on the current value of $x$.
The counterpart \verb|!id:(x := f);| does not exist.
The example below amounts to showing $x + 1 > x$ for all $x$.

\begin{lstlisting}[language=kaisar]
x := *; y := x + 1; ?zFact:(z := y);
!compare:(z > x) using zFact ... by auto;
\end{lstlisting}

\subsection{Verifying Demonic and Angelic loops}
\label{sec:kaisar-for-loops}
We add Demonic and Angelic loops.
Demonic loops, where Demon chooses when to repeat the loop, are written  \verb|{<body>}*|.
Angel proves a Demonic loop using an invariant, which she usually gives a name, e.g., \verb|inv|.
Identifying a loop invariant is a key proof step where human insight is required to overcome the undecidability of hybrid systems.
First, Angel proves the base case just before the loop.
In the body, fact name \verb|inv| stands for the assumption that the invariant held at the body's start.
The body must end with a proof of the same invariant in the body's final state.
Used after the loop's end, \verb|inv| is the now-proven fact that the invariant holds at the loop's end.
The below example shows $x \geq y$ is an invariant if $x$ initially equals a constant $y,$ then increases.

\begin{lstlisting}[language=kaisar]
?yZero:(y := 0); ?xZero:(x := 0); ?cPos:(c = 3);
!inv:(x >= 0);
{ x := x+c; !inductiveStep:(x >= 0) using cPos inv by auto; }*
!geq:(x >= y) using inv yZero by auto;
\end{lstlisting}

We also take this opportunity to review  Kaisar's persistent fact naming: fact \texttt{xZero} stays accessible even after the loop modifies \texttt{x}, but merely says the \emph{old} value of $x$ was zero, as it would be unsound to assume that $x$ remained zero in or after the loop. At the end, clearly \texttt{yZero} and \texttt{cPos} still hold.

Angelic loop proofs are based on \texttt{for} loops, but differ by having additional notation for invariant and termination reasoning.
Our toy example will compute the triangular numbers by sums and prove that Gauss's formula is their solution.
Non-toy uses of \texttt{for} loops in CPSs are in \rref{sec:kaisar-idioms}.

We first discuss loop headers, which have four parts.
We present and discuss the header for our summation example here, then give the full example after discussing the header.
\begin{lstlisting}[language=kaisar]
 for (x := 1; !(sum = sol(x)); ?(x <= 11); x := x + 1) { ...... }
\end{lstlisting}

We describe each part and its role in the following example:
\begin{inparaenum}[\it i)]
\item An assignment initializes the loop index variable (here, \texttt{x} is initialized to $1$).
\item An invariant is proved to hold initially (here, \texttt{sum = sol(x)} solves the \texttt{sum} as a function of \texttt{x}).
When the invariant fact is mentioned in the loop body, it means the invariant held at the start of the body.
\item A guard condition is provided which determines when the loop stops.  
To ensure termination, it must contain (perhaps as a conjunct) an inequality which gives a locally-constant upper (respectively lower) bound on an increasing (respectively decreasing) index variable.
In the example, \texttt{x <= 11} bounds \texttt{x} above by 11, ensuring termination.
The value 11 is an arbitrary example.
Advanced uses of guards often combine upper bounds (which show termination) and lower bounds (which show progress).
\item
The index is modified (here, incremented by \texttt{1}).
The sign and constancy of the increment (1) are checked; because 1 is positive and the guard bounds \texttt{x} above, the loop is proved to terminate.
\end{inparaenum}

\begin{lstlisting}[language=kaisar]
 ?deltaLo:(delta > 0);
 ?deltaHi:(delta < 1);
 let sol(x) = x*(x+1)/2;
 sum := 1;
 for (x := 1; !(sum = sol(x)); ?(x <= 11); x := x + 1) {
    sum := sum + (x+1);
   !step:(sum = sol(x+1));
 }
 !done:(x >= 11 - delta) by guard(delta);
 !total:(sum >= 50) using done sum x deltaHi by auto;
\end{lstlisting}
Above, we first assume that \texttt{delta}, which we use for sound real-number comparisons, is in the interval $[0,1]$.
Next, we \texttt{sol}ve the $x$'th triangular number in terms of $x$ by Gauss's formula.
Variable \texttt{sum} is initialized to the first triangular number (1) and stores the $x$'th triangular number as $x$ grows.
We recall the loop header's meaning: $x$ ranges from 1 to 11 (for example) and Gauss's formula is the invariant.
The loop body adds $x+1$ to \texttt{sum}, thus \texttt{sum} becomes the next ($x+1$) triangular number.
The next \texttt{step} asserts that Gauss's formula holds for $x+1$ assuming it holds for $x$.

Constructivity makes Kaisar's real arithmetic subtle: exact real comparisons are undecidable, so constructive comparisons are inexact.
The user optionally specifies the loop guard's comparison precision (above: \texttt{delta}) with the special \texttt{guard} proof method.
We learn \texttt{x >= 11 - delta} upon termination since loops end once Kaisar cannot \emph{prove} the guard true;
in the worst case, the guard holds, but only by a small (\texttt{delta}) margin.
The assertion \texttt{total}  derives a final bound on \texttt{sum} from the bound on \texttt{x}.
Note that comparison inexactness is one-sided: termination consequence \texttt{done} is made inexact by a margin of \texttt{delta} so that loop guard \texttt{x <= 11} need not be made inexact.

In serious uses (\rref{sec:kaisar-idioms}), the compared quantities will be proper real numbers rather than natural numbers, so that inexact comparisons become essential to ensure constructivity and thus essential to ensure that proofs correspond to executable code. The above example may surprise the reader in its use of real numbers to model basic properties of natural numbers; this use is atypical in practice and serves only as a simple demonstration of \texttt{for} loops.

The argument of the \texttt{guard} method can be omitted. If so, \texttt{guard} heuristically searches the context for positive constants which it reuses in an effort to reduce the number of variables used.


\subsection{Verifying Hybrid Games}
\label{sec:kaisar-hybrid-games}
We now make game strategies \emph{hybrid} by adding ordinary differential equation proofs (ODE proofs), then adding crucial proof principles including  differential induction, cuts, and solutions~\cite{DBLP:journals/jar/Platzer17,Platzer18}.
This combination of proof principles reflects Kaisar's goal of making easy proofs easy but hard proofs possible: solution reasoning improves automation for simple ODEs while differential induction reasoning supports polynomial ODEs which need not have any closed-form solution, by exploiting human insights in the form of invariants.
In \CdGL, the equations of an ODE system are comma-separated, followed optionally by a \emph{domain constraint} specifying formulas which must hold throughout the duration of the ODE.
Each domain constraint element is prefixed by the \verb|&| symbol like a conjunction.
In \CdGL, the current player is responsible for choosing the ODE duration and proving the domain constraint holds at all times up to and including the duration.
The syntax of an ODE \emph{proof} in Kaisar generalizes the syntax of an ODE system:
the statements in a \emph{domain constraint proof} include assumptions  (\verb|?|) and/or assertions (\verb|!|), which are respectively
assumed or proved to be true at all times throughout the ODE's evolution.
To make their syntax visually distinct, ODE system proofs are wrapped in braces before their terminating semicolon. We demonstrate the major proof principles on Demonic ODEs first, then add support for Angelic ODEs.

Assertions (\verb|!|) in Demonic ODE proofs correspond to \emph{differential cuts}~\cite[Ch.\ 11.11]{Platzer18} in \CdGL, meaning they are proved rather than assumed.
Differential cuts are crucial in both theory and practice because, in combination with differential induction, they can prove facts that the latter could not prove alone~\cite[Ch.\ 11.11]{Platzer18}. To prove that a differential cut holds at all times throughout an ODE, Kaisar fixes a time, assumes that previous assertions (and all assumptions) in the domain constraint hold at that time, then  proves the cut (assertion) formula holds at that time.
An assertion in an ODE \verb|auto|matically reasons by its solution if available, else by induction.
The \verb|solution| and \verb|induction| methods force the respective approaches. The \verb|solution| method reports an error on an ODE whose solution is non-polynomial.

As the next example shows, for ODEs whose solutions are polynomial Kaisar terms, streamlined solution reasoning is available because polynomial solutions are amenable to automatic arithmetic solving.
The assertion checker automatically uses solutions when relevant.
For manual control, explicit assertions (\verb|xSolAgain| below) can also be used.
Recall (\rref{sec:kaisar-discrete}) that  \verb|?xInit:(x := 2);| 
binds \verb|xInit| to the equality induced by the initializing assignment \verb|x := 2|.
\begin{lstlisting}[language=kaisar]
?xInit:(x := 2); y := 0;
{y' = 1, xSol: x' = -2 & ?dc:(x >= 0) & !xSolAgain:(x = 2*(1 - y))};
!xHi:(x <= 2) using xInit xSol by auto; !xLo:(x >= 0) using dc by auto;
\end{lstlisting}

\emph{Differential induction}~\cite[Ch.\ 10-11]{Platzer18} is crucial because real  models often use ODEs that do not have closed-form solutions, let alone polynomial ones. In contrast to the solvable fragment above, Kaisar's  differential induction allows verifying ODEs with polynomial \emph{right-hand sides}, a broad class containing both linear and non-linear equations.
Differential induction can support this broad class of ODEs because it reasons analytically on the \emph{definition} of the ODE and need not know its \emph{solution}. For example, an equality $f = g$ between differentiable terms holds by induction if $f=g$ holds initially and $\der{f} = \der{g}$ holds throughout. This argument does not rely on ODE-solving.

Differential induction proofs have base cases and inductive cases like loops do, but differ in that it is optional to write base cases explicitly.
Kaisar attempts an \verb|auto|matic base case proof by default.
To use an explicit proof instead, assert the base case just before the ODE.

Our next example, circular motion, is a common use of differential induction, since many circular models are nonlinear (e.g., multi-affine~\cite{DBLP:journals/ral/BohrerTMSP19}) and even simple ones have undecidable trigonometric solutions. Our example models constant-speed rotation and proves  (\verb|circle|) that $(x,y)$ stays on the unit circle.  In \rref{sec:kaisar-evaluation}, the same differential induction rule is used in serious case studies.

\begin{lstlisting}[language=kaisar]
x := 0; y := 1; {x' = y, y' = -x & !circle:(x^2 + y^2 = 1) by induction};
\end{lstlisting}

Next we discuss \emph{Angelic} ODEs, where Angel, not Demon, chooses the ODE duration
and must prove the domain constraint. 
To make an ODE Angelic, we add one assignment to the domain constraint to specify Angel's chosen duration. The assigned variable, often named $t$, must be a clock, i.e., have initial value \verb|t:=0| and derivative \verb|t'=1|. Because Angel is responsible for proving the domain constraint, assumption statements are disallowed in every Angelic domain constraint proof.
The below example models 1D driving with an Angelic ODE.

\begin{lstlisting}[language=kaisar]
?(T > 0); ?accel:(acc > 0);
x := 0; v := 0; t := 0;
{t' = 1, x' = v, v' = acc 
 & !vel:(v >= 0) using accel by induction
 & !vSol:(v = t * acc) by solution
 & !xSol:(x = acc*(t^2)/2) by induction
 & ?dur:(t := T)};
!finalV:(x = acc*(T^2)/2) using dur xSol by auto;
\end{lstlisting}
The reification of ODE proofs into  \CdGL games (\rref{sec:kaisar-relationships}), differs subtly between Demonic and Angelic ODEs. In the Demonic case, the reified domain constraint contains only the assumptions, because Demon is responsible for showing the domain constraint of a Demonic ODE, which Angel can thus assume.
The reified domain constraint of an Angelic ODE contains assertions, since Angel is responsible for proving the domain constraint;
indeed, assumption statements are prohibited in Angelic domain constraint \emph{proofs}.
If the user wants to exclude an assertion from an Angelic ODE's reification, they enclose it in a (forward) ghost, which we describe next.
Ghosts are not specific to ODEs; they can be applied to any sequence of Kaisar statements.

\subsection{Uniform Ghost Reasoning}
\label{sec:kaisar-ghost-proof}
Recall that it is important for Kaisar to check (\rref{sec:kaisar-relationships}) \emph{what} theorem is proved by each strategy in order to clearly communicate the results of verification, but the checker sometimes requires hints from the user.
Those hints are crucially provided using Kaisar's notions of forward and inverse \emph{ghost statements}, which are respectively written \verb|/++ pf ++/| and \verb|/-- pf --/|. Informally, a forward ghost \verb|/++ pf ++/| belongs to the proof but not the game, while inverse ghost \verb|/-- pf --/| belongs to the game but is unused in the proof. 
By putting different statements inside ghosts, we provide uniform notation for several standard logical rules.
The \emph{weakening} and \emph{differential weakening} rules, which respectively hide facts and ODE dynamics, respectively amount to inverse-ghosts of facts and domain constraint statements.
The \emph{cut} and \emph{differential cut} rules, which respectively prove lemmas for use as facts and domain constraint elements, amount to forward-ghosts of facts and domain constraint assertions.
Forward ghosts of assertions of lemmas are particularly common in practice because proofs often feature lemmas not shown in the model.
Forward \emph{differential} ghosts also enable otherwise-impossible proofs~\cite[Ch.\ 12]{Platzer18}, such as proofs of exponential decay properties. Inverse ghosts are used less frequently, but make the language design more symmetric and can assist proof automation by signaling that ghosted facts should not be selected as assumptions.

The proof-checker enforces sound scoping rules for ghosts.
For any $y$ bound by a forward-ghost assignment $\humod{y}{f},$ free dependencies on $y$ can only appear in other forward ghosts: it would be unsound to continue assuming $y=f$ after erasing the assignment $\humod{y}{f}$. Facts introduced within an \emph{inverse} ghost can only serve as assumptions to other inverse-ghost proofs, because inverse-ghosting of facts represents a weakening proof step, which hides or ignores a fact.
In contrast, non-ghost proofs may use forward-ghost facts because erasing an assertion makes its conclusion no less true. 
Subtly, facts $p(y)$ about a forward-ghost variable $y$ can be used to prove non-ghost facts $q()$ which do not mention $y$.
Such proof steps remain sound after ghost erasure because, in contrast to facts mentioning $y,$ they say mere \emph{existence} of $y$ satisfying $p(y)$ implies $q(),$ an argument which holds equally well if $y$ is never actually assigned. Demonic ODEs' domain constraints treat their assertions like forward ghosts because only their \emph{assumptions} correspond to the domain constraint of a \CdGL game, but Angelic ODEs do not, because their \emph{assertions} do correspond to the \CdGL game's domain constraint.
Crucially, Kaisar enforces these soundness rules automatically to protect the user from accidentally-unsound uses of refinements.

The next example shows \emph{forward}-ghost  assertions and assignments.
The example  remembers the value of $x$ in $y$ with a ghost assignment.
Ghost assignments will be essential in \emph{differential} ghost proofs to initialize ghost variables before an ODE.
At the same time, the example is preparation for line labels (\rref{sec:nominals}), which provide a more automatic mechanism to remember states in practice.

\begin{lstlisting}[language=kaisar]
?xInit:(x > 0);
/++ ?yInit:(y := x); !inv:(x >= y);  ++/
{ x := x + 1; /++ !(x >= y) using inv by auto; ++/ }*
!positive:(x > 0) using inv yInit xInit by auto;
\end{lstlisting}
Assertions such as \verb|inv| which mention $y$ were crucially ghosted for soundness, whereas assertion \verb|positive| does not mention ghost $y$ and need not be ghosted.

Next, we use a \emph{differential ghost} to add a continuously-changing variable $y$ to an ODE.
We show a canonical use: differential ghosts are essential for proving invariants of exponentially-decaying systems because those properties are often non-inductive: they approach falsehood without reaching it. To use a differential ghost variable $y,$ we  initialize $y$ with a ghost assignment, then give a ghost proof of an \verb|inv|ariant. The invariant is proved true at all times throughout the ODE, after which it is used to prove a conclusion (e.g., \verb|positive|) which does not mention the ghost.

\begin{lstlisting}[language=kaisar]
x := 1; /++ y := (1/x)^(1/2); !inv:(x*y^2 = 1) by auto; ++/
{x' = -x, /++ y' = y * (1/2) ++/ & !inv:(x*y^2 = 1) by induction};
!positive:(x > 0) using inv by auto;
\end{lstlisting}

Soundness requires that the addition of a forward differential ghost variable does not reduce the ODE's existence interval (by introducing an infinite asymptote~\cite[Ex.\ 12.2]{Platzer18}). To ensure this, it suffices~\cite[Lem.\ 12.2]{Platzer18} that Kaisar require that the right-hand side for $\D{y}$ is linear in $y$. The invariant for the ghosted system (here, $x \cdot y^2 = 1$) sometimes seem unintuitive, but 
can be constructed systematically~\cite[Ch.\ 12]{Platzer18} for all true differential equation invariants \cite{DBLP:journals/jacm/PlatzerT20}. Kaisar's enforcement of these soundness conditions crucially protects the user should they attempt an unsound ghost proof.

We now discuss \emph{inverse} ghosts.
Whereas forward ghosts represent proof statements not appearing in a game, inverse ghosts  represent elements of a game which must not be used in the proof.
Inverse ghosts in ODEs can be used to forget irrelevant dynamics in order to optimize (automatic or manual) reasoning about the remaining dimensions of the dynamics.
Below, we forget unsolvable dimensions which move in a circle and reason by the (linear) solution of the other dimension $z$.
\begin{lstlisting}[language=kaisar]
z := 0;  {/-- x' = y, y' = -x --/,  z'=1 & !zPos:(z >= 0) by solution}
\end{lstlisting}

Inverse ghost \emph{tests} can be useful to indicate that a test should not be selected by automated fact selection heuristics in following proof steps.
The toy example below supposes we wish to model a 3-dimensional system where motion occurs in only one dimension $x$.
Inverse ghosts provide machine-checked documentation that assumptions on the dimensions other than $x$ are superfluous for the proof and can thus be safely erased.

\begin{lstlisting}[language=kaisar]
x := 0; /-- y := 25; z := -10; --/ {x' = 3} !(x >= 0);
\end{lstlisting}

\subsection{Time-Traveling Proofs with Labeled Reasoning}
\label{sec:nominals}
We introduce labeled reasoning, a Kaisar feature which greatly generalizes previous principles for references across states (\rref{sec:kaisar-relwork}) by freely mixing past, future, and hypothetical states.
Labeled reasoning is applied in \rref{sec:kaisar-idioms}.

Specifically, we allow a line (location) to be given a \verb|label:| and allow statements elsewhere to write \verb|expr@label| to mean the value of the \verb|expr|ession as of the \verb|label|ed line.
The \verb|label:| syntax is an allusion to labels in assembly code.
In stark contrast to assembly programs, Kaisar proofs \emph{raise} their abstraction level by using labels, which free the user from manually determining the value of an expression at a state and manually remembering it in ghost variables.
As defined elsewhere~\cite[7.3.6]{Bohrer21}, we use an analog of \emph{static single assignment} (SSA) for systematic variable numbering in proofs, under-the-hood.
We informally discuss SSA throughout the labeling discussion.
In contrast to the use of SSA in compilers, our notion of SSA for proofs serves as a consistent variable naming system so that high-level proof techniques for past and future states can be implemented systematically.

\paragraph{Terminology.}
We briefly define core concepts.
The statement \verb|label:| is referred to as a label statement.
We call its location the \emph{labeled point}.
Informally, we refer to the point by the name of the label.
The expression \verb|e@label| is a \emph{located expression} of  expression \verb|e| located at \verb|label|.
Point \verb|label| is the \emph{referent} of the located expression while the point at which \verb|e@label| is written is called the \emph{referrer}.
The \emph{difference} between any two points, when it exists, is the list of statements which are passed through on \emph{every} path from the point that appears earlier to the one that appears later.
Kaisar features an SSA translation where a single variable $x$ from the Kaisar source is translated to a family of variables 
(SSA-variants) $x_i,$ where $x_0$ is the initial value of $x$.
An expression $e$ is \emph{mobile to} \verb|label| if every free (subscripted) variable $x_i$ of $e$ for which $i > 0$ has been assigned by point \verb|label|, e.g., $x_0$ is mobile to everywhere.
The subscripted variables $x_i$ are crucial for making mobility possible because they remember old values of $x$ throughout the future even if the current value of $x$ gets 
overwritten\footnote{
A proof following a loop will not remember each of the loop's intermediate values because there are finitely many $x_i$.
If $x_i$ is bound in one iteration, it can be overwritten in the next, and again after the loop ends.
}.
The \emph{current} variant $x_i$ of $x$ at \verb|label| is the $x_i$ with the greatest $i$ which is mobile to \verb|label|.
We write \verb|x@label| interchangeably with the current $x_i$ at \verb|label| when doing so promotes readability.
When Kaisar knows the solution and duration of an ODE, the discussion of assignments extends to ODE solutions.

\subsubsection{Historical Proof with Backward Labels}
Backward label references are the simplest and are commonly used to remember initial states of loops, ODEs, or the overall strategy.
Kaisar remembers for each $x$ and \verb|label| which $x_i$ was current at \verb|label|.
A backward reference \verb|expr@label| simply replaces each free variable $x$ of \verb|expr| with the $x_i$ associated to \verb|label|.
Below,  \verb|x@init| becomes $x_0$:
\begin{lstlisting}[language=kaisar]
init: ?(y = 0); !bc:(y = 2*(x - x@init));
{ x := x + 1; y := y + 2; !step:(y = 2*(x - x@init)); }*
\end{lstlisting}
ODE proofs often inductively track the change in a quantity; in the below example, it increases:
\begin{lstlisting}[language=kaisar]
old: {x' = 1 & !greater:(x >= x@old)};
\end{lstlisting}
or track the fact that a quantity, such as $x+y$ below, is a conserved quantity:
\begin{lstlisting}[language=kaisar]
x:=0; y:=0; start:
{x' = 1, y' = -1 & !conserved:(x+y = (x+y)@start)};
\end{lstlisting}

\subsubsection{Predictable Futures and Forward Determined Labels}
\emph{Forward} references are a powerful application of labels, yet surprising because the future is often unpredictable.
To ease the introduction of forward references, we first discuss the special case of forward \emph{determined} references with predictable futures.
A forward reference is \emph{determined} if the difference (path) between referrer and referent contains only deterministic assignments, Angelic or Demonic tests, and non-program Kaisar statements.
This fragment is a natural stepping stone because the change in state between referrer and referent can be expressed entirely with deterministic assignments.
To resolve \verb|e@label| below, start with \verb|e| and apply each assignment in the difference as a substitution, in reverse order.
\begin{lstlisting}[language=kaisar]
x := 0;
init: !(x < x@final); x := x + 1; x := x + 2;
final:
\end{lstlisting}
The difference between \verb|init| and \verb|final| above contains \verb|x := x + 1;| and
 \verb|x := x + 2;| so that \verb|x@final| resolves to \verb|(x@init + 1) + 2|.
Crucially, a resolved expression is always mobile to the referrer (e.g., \verb|init|), i.e., an SSA-variant $x_i$ is never mentioned before assignment.
Next, differences of determined labels cannot contain choice (\verb|++|) statements, but may cross choice boundaries:
\begin{lstlisting}[language=kaisar]
x := 0; y := x@mid;
init: { {x := x + 3; mid: x := x * x;} ++ x := 5; }
\end{lstlisting}
When $y$ is assigned, we do not know whether \verb|mid| will be evaluated, but know that \emph{if} \verb|mid| is reached, $x=3$ holds there.
Thus, we statically elaborate \verb|x@mid| to \verb|0 + 3| so that $y$ will have value 3.
Even the parallel branch \verb|x := 5| is permitted to use \verb|x@mid| with value 3, but such references are rarely useful.
References which \emph{exit} choices are also supported; the following example also gives $y$ value 3.
\begin{lstlisting}[language=kaisar]
{ {y := x@final; x := 2;} ++ x := 5;} x := x + 1;
final:
\end{lstlisting}
The approach discussed thus far does not support two important cases: the cases of cyclic dependencies and 
references across nondeterministic differences.
Cyclic dependencies are fundamentally an error and are reported by Kaisar automatically.
An example is:
\begin{lstlisting}[language=kaisar]
x := x@two; one: x := x@one; two: 
\end{lstlisting}
Each assignment asks the other what value to give $x$. This cycle is an error because it leaves both values of $x$ undefined.
In contrast, references across nondeterministic differences  are not fundamentally errors, 
they simply require a generalization which we present below.

\subsubsection{Unpredictable Futures and Hypothetical Label Reasoning}
\label{sec:kaisar-forward-label-param}
When a forward located expression's value is not unique, we reason \emph{hypothetically}: ``Suppose $x$ gets assigned $y$, what is the value of $e$?''
We extend our prior notation with arguments: label statement \verb|label(var1, ..., varN):| lets program variables \verb|var1, ..., varN| be replaced with arguments \verb|f1, ..., fN|.
Located expression \verb|e@label(f1, ... fN)|, now replaces each \verb|varI| with hypothetical value \verb|fI| during resolution (after resolving each \verb|fI| recursively if needed).
To support forward references between arbitrary locations, one need only introduce arguments for any variables which were bound nondeterministically along the difference; Kaisar will report the variables that must be made arguments, if any.
The rest of the section implements major CPS proof idioms with Kaisar's hypothetical labels.

\subsection{Proof Patterns for CPS}
\label{sec:kaisar-idioms}
We now demonstrate, by example, how Kaisar in general and labeled reasoning in particular streamline proofs of major recurring CPS proof idioms. Our first idiom is \emph{logical model-predictive control}~\cite[Sec.\ 8.1]{Loos16}, a proof idiom  inspired by \emph{model-predictive control}~\cite{DBLP:journals/ijrr/MitschGVP17}.
This idiom predicts the physical change resulting from each available control choice to determine a range (or envelope) of safe decisions.
The \emph{sandbox} paradigm extends the approach by Demonically assigning a control choice and checking it against the model.
The sandbox paradigm is crucial for combining verified models with practical implementations (\rref{sec:kaisar-future-work}) because it treats (potentially-complex) controller implementations as untrusted black-boxes. 
Sandboxing has  been used in case studies because of these practical implications (\rref{sec:kaisar-evaluation}).
We give a 1D driving example where Angel wants to  keep a safe braking distance $\textit{SB}()$ between her and the goal $(d - x)$
while Demon controls the ODE duration $t \in [0, T]$.
Angel predicts the effect of each acceleration  in the worst case $t = T$ and allows (\texttt{env}) all accelerations that preserve $\textit{SB}() \leq d - x$.
The proof completes by an arithmetic step showing that safety for $t = T$ implies safety for all $t \in [0, T]$.
Recall from \rref{sec:kaisar-prop-proofs} that nullary syntax just means function \verb|SB()| has no parameters;
it may still depend on the state.

\begin{lstlisting}[language=kaisar]
let SB() = v^2/(2*B); let safe() <-> (SB() <= (d-x));
?pre:(T > 0 & A > 0 & B > 0); ?initSafe:(safe());
{acc := *; ?env:(-B <= acc & acc <= A & safe()@ode(T));
 t := 0; {t'=1, x'=v, v'=acc  & ?time:(t <= T) & ?vel:(v >= 0)};
ode(t): !step:(safe()) using env pre time vel ... by auto;
}*
\end{lstlisting}
In step \verb|env| above, Kaisar successfully automated the (non-trivial) resolution of the high-level located expression \verb|safe()@ode(T)|.
Symbol \verb|safe()| depends on \verb|SB()|, which depends on \verb|v| and \verb|x|.
Kaisar automatically determined the values at location \verb|ode| (with $t=T$) by solving the ODE:
\[ 
x@ode(T) = x + v \cdot T + \texttt{acc} \cdot T^2/2 \quad\quad
v@ode(T) = v + \textit{acc} \cdot T \quad\quad
t@ode(T) = T
\]
Thus \verb|safe()@ode(T)| resolves to $(v + \text{acc} \cdot T)^2/(2 \cdot B) \leq d - (x + v \cdot T + \text{acc} \cdot T^2/2)$.
For ODEs whose solutions Kaisar cannot compute, $x$ and $v$ would be made into label parameters.

Shown below, a major way that hypothetical references streamline idioms including logical model-predictive control is that they automate deriving the meaning of a high-level specification like \verb|safe()@ode(T)|.
Even \verb|SB()|  can be derived from its first principles:  \verb|SB()| is the least distance \verb|d-x| for which full braking \verb|B| preserves \verb|safe|ty throughout the time \verb|ST()| where the vehicle fully stops. 
We then define \verb|safe|ty as obeying \verb|x<=d| at the stopping time \verb|ode(ST())|, i.e., the final state of the ODE whose duration was \verb|ST()|.
Lastly, \verb|print(safe())| prints the text of \verb|safe()| to the user, who uses it to define \verb|SB()|.
The first-principles derivation of \verb|ST()| is analogous and omitted.

\begin{lstlisting}[language=kaisar]
?(B > 0); 
let ST() = v / B; 
!stopTime:(v@ode(ST()) = 0);
let safe() <-> x@ode(ST()) <= d;
t := 0; {t' = 1, x' = v, v' = -B  & ?(v >= 0)};
ode(t): print(safe());
\end{lstlisting}
For this solvable ODE, \verb|safe()| resolves to $x  + v \cdot (v / B) + \frac{-B \cdot (v / B)^2}{2} \leq d,$ yielding \verb|SB() = v^2/(2*B)|.

Next, we extend the predictive model with a \emph{sandbox controller}.
The \verb|switch| statement below implements the sandbox, with the fallback guarded by \verb|true| since it is provably safe regardless of \verb|accCand|.
The guard \verb|true| makes Kaisar's \verb|case| totality check succeed trivially.
Each assertion \verb|predictSafe| reasons predictively.
The latter assertion predicts motion at time \verb|min(T,v/B)| specifically to capture the case where the system brakes to a complete stop early, i.e.,  $t < T$.
The assertion \verb|!step| uses disjunctive lookups to be concise: the disjunction of the \verb|predictSafe| assertions implies safety in the worst case, which (by arithmetic) implies safety in every case.

\begin{lstlisting}[language=kaisar]
let SB() = v^2/(2*B);
let safe() <-> SB() <= (d-x) & v >= 0;
?pre:(T > 0 & A > 0 & B > 0);
?initSafe:(safe());
{ accCand := *;
  let admiss() <-> -B <= accCand & accCand <= A;
  let env()    <->  safe()@ode(T, accCand);
  switch {
    case inEnv:(env()) =>
      ?theAcc:(acc := accCand);
      !predictSafe:(safe()@ode(T, acc));
    case true =>
      ?theAcc:(acc := -B);
      !predictSafe:(safe()@ode(min(T,v/B), acc));
  }
  t:= 0;
  {t' = 1, x' = v, v' = acc & ?time:(0 <= t & t <= T) & ?vel:(v >= 0)};
ode(t, acc):
  !step:(safe()) using predictSafe pre initSafe time vel ... by auto;
}*
\end{lstlisting}
Because Kaisar's underlying logic \CdGL is constructive, we briefly discuss the computational interpretation of \verb|switch| above.
The \verb|switch| is executed by testing whether \verb|env()| holds and applying the \verb|true| branch only when \verb|env()| cannot be shown true.
The \verb|env()| branch is first (because its guard allows accelerating) so taken whenever possible, whereas \verb|true| branches must be conservative, e.g. by braking.
\CdGL compares numbers up to some precision $\delta > 0$ because exact equality is undecidable for its real numbers.
The \verb|env()| branch is always taken when \verb|env()| holds by a margin of $\delta,$ but either branch can be taken when \verb|env()| holds by a margin in $[0, \delta)$.

Precisions $\delta > 0$  are inferred by Kaisar.
For \texttt{switch}es with \texttt{true} branches (i.e., the one above) \emph{any} $\delta > 0$ suffices.
Recall that Kaisar checks constructively the validity of the disjunction of guards; precision $\delta$ is inferred during this check, 
e.g., disjunction $x \geq y \lor x \leq y + \delta$ for $\delta > 0$ is valid with margin $\delta,$ but disjunction $x \geq y \lor x < y$ 
is not constructively valid since comparison is inexact.
See \rref{sec:kaisar-for-loops} for comparison $\delta$s in the context of loop guards.
Related work on \CdGL~\cite{DBLP:conf/cade/BohrerP20} describes the foundations of inexact comparisons.
In short, the constructive foundations are type theory and constructive real analysis, which dictate that inexact comparisons, but not exact ones, are decidable and thus constructively valid.
Though constructive validity is the core truth notion of Kaisar and \CdGL, robust truth helps explain this notion by analogy.
If comparison $f \geq g$ holds robustly by a margin of at least $\delta > 0,$ 
Kaisar can always prove $f \geq g$ constructively. 
If $f \geq g$ holds  by a margin less than $\delta,$ Kaisar either reports $f \geq g$ or $f \leq g + \delta,$ each true by low margins.

Next, we conclude the series of examples by introducing the reach-avoid proof paradigm.
This paradigm is important because it shows two fundamental correctness properties together: safety and liveness.
We model a 1-dimensional, velocity-controlled vehicle which stops as close as it can to some goal \verb|d|.
Reach-avoid proofs are an important CPS application of \texttt{for} loops (\rref{sec:kaisar-for-loops}).

\begin{lstlisting}[language=kaisar]
?epsPos:(eps > 0); ?consts:(x = 0 & T > 0 & d > eps);
init:
for (pos := 0;
    !conv:(pos <= (x-x@init) & x <= d) using epsPos consts ... by auto;
    ?guard:(pos <= d - (eps + x@init) & x <= d - eps);
    pos := pos + eps/2) {
  vel := (d - x)/T;  t := 0;
  {t' = 1, x' = vel & ?time:(t <= T)};
  !safe:(x <= d) using conv guard vel time by auto;
  ?high:(t >= T/2);
  !prog:(pos + eps/2 <= (x - x@init)) using high ... by auto;
  note step = andI(prog, safe);
}
!done:(pos >= d - (eps + x@init) - eps | x >= d - eps - eps) by guard;
!(x <= d & x + 2 * eps >= d) using done conv by auto;
\end{lstlisting}
The above model starts by assuming bounds on constants and the state.
The initial state is labeled \texttt{init}.
The loop header initializes the index variable \texttt{pos},
then bounds it above by the change in
\texttt{x} while position \texttt{x} is bounded above by goal  \texttt{d}.
To ensure termination, the guard bounds \texttt{pos} above by a (locally) constant term based on $x$'s initial value.
The guard also requires at least \texttt{eps} remaining distance to the goal, which will help us show progress.
The factor of \texttt{1/2} in the final header statement reflects the subtle adversarial nature of games: 
Demon can pick any duration $t \in [T/2, T],$ so \texttt{eps/2} reflects the motion in the worst case $t = T/2$.

The proofs of safety and liveness, more broadly, demonstrate the rich back-and-forth dynamics in games.
First, Angel makes a control choice for \verb|vel| without knowing the ODE duration $t$.
Angel sets \verb|vel| to \verb|(d-x)/T| so that the goal is reached in the best case and at least \verb|eps/2| progress is made in the worst case.
Next, Demon chooses the duration of the ODE, which is equal to the value of $t$ upon termination of the ODE.
Angel proves safety under weak assumptions on timing: $0 \leq t \leq T$.
It is essential that Angel proves safety without assuming a positive lower bound on $t,$ as safety ought to hold at all times.
However, liveness needs a positive lower bound: if the ODE were to evolve for 0 time, no progress would be made.
Thus, Demon announces a lower bound $t \geq T/2,$ but it only becomes available to Angel as an assumption \emph{after} safety is proved.
The specific lower-bound $T/2$ is chosen for the sake of simplicity.
Angel is then responsible for proving the  invariant, which includes both progress (\verb|prog|) and safety (\verb|safe|) invariants in a reach-avoid proof.
Specifically, it is proved that invariants \verb|prog| and \verb|safe| will hold \emph{after} the update
\verb|pos := pos + eps/2| is executed, thus the assertion \verb|prog|  writes \verb|pos + eps/2| where
\verb|conv| wrote \verb|pos|.
The proof of \verb|safe|ty appeals to \verb|conv|, which, when accessed from the loop body, supplies the fact that the invariant held at the beginning of the body.
The \verb|note| statement conjoins \verb|prog| and \verb|safe| to show the full invariant.

In the above example, the optional argument of the \texttt{guard} method is omitted, thus \texttt{guard} heuristically chooses a comparison $\delta$.
We review the role of inexactness resulting from both \verb|eps| and constructivity.
Recall that comparisons are inexact: the \verb|guard| tactic optionally accepts a comparison $\delta$ as an argument,  else $\delta$ is chosen heuristically.
In this case, $\delta=\text{\texttt{eps}}$ is chosen automatically, but constructive comparison is not the primary purpose of \verb|eps|.
Rather, the use of \verb|eps| in the guard ensures the body only runs when the distance remaining is bounded below, letting us prove a lower bound on progress and thus prove we reach the goal.
Inexact comparisons make their impact in fact \verb|done|, where $\delta=\text{\texttt{eps}}$ is subtracted from the guard term to account for uncertainty.
Note that only one branch of the comparison is made more inexact, specifically \verb|done|.

This concludes our presentation of Kaisar by example, the first proof-checking language and tool for \CdGL.
Labeled reasoning was a major novel feature which simplified key proof paradigms for hybrid games.
At the same time, Kaisar's design combined high-level features including definitions, proof terms, ghosts, and lexical scope in the challenging context of proving hybrid games.

\section{Evaluation via Case Studies}
\label{sec:kaisar-evaluation}
We evaluate Kaisar against Bellerophon~\cite{DBLP:conf/itp/FultonMBP17}, the unstructured proof language of \KeYmaeraX, a theorem prover for (classical) hybrid systems and games in \dL and \dGL. 
We ported three driving case studies from the literature (PLDI-DC~\cite{DBLP:conf/pldi/BohrerTMMP18}, IJRR~\cite[Thm.\ 1]{DBLP:journals/ijrr/MitschGVP17}, RA-L~\cite[Thm.\ 1]{DBLP:journals/ral/BohrerTMSP19}) from Bellerophon to Kaisar. We then generalized PLDI-DC in four stages (PLDI-AS, PLDI-TAC, PLDI-RA, PLDI-RAD), which we back-ported to Bellerophon for an additional comparison. We record metrics (\rref{sec:kaisar-evaluation-stats}) about these case studies which, together with the examples from previous sections, serve as proxies that help assess Kaisar's core goals such as its  productivity, ability to verify non-trivial systems, and maintainability. The IJRR and RA-L models explore Kaisar's capabilities when applied to larger models, while the PLDI series of models demonstrates Kaisar's ability to co-evolve models and proofs over time.
We first present one case study in both the Bellerophon language and Kaisar for comparison in \rref{sec:kaisar-evaluation-example}, then present detailed statistics and discussion for all case studies in \rref{sec:kaisar-evaluation-stats}.

\subsection{Side-by-side Comparison of Kaisar and Bellerophon}
\label{sec:kaisar-evaluation-example}
Figure~\ref{fig:pldi-tac-listings} shows the PLDI-TAC example in both Kaisar and Bellerophon.
This example was chosen to demonstrate both Kaisar's Angelic strategy connectives and adversarial timing.
Our discussion is high-level; detailed descriptions of Bellerophon are in the literature~\cite{DBLP:conf/itp/FultonMBP17}.

The Kaisar version of PLDI-TAC 
is much more concise, both due to combining definitions, models, and proofs in one artifact and due to proof automation.
All assertions (\texttt{!}) in the Kaisar model proved without manual help.
Bellerophon has much proof automation, but little for Angelic loops.
It must use the \texttt{con} tactic to manually write a variant predicate that holds in each iteration as its fresh argument $k$ decreases.
The Bellerophon proof also unfolds program statements and proves the ODE with differential cuts (\texttt{dC}).
The Bellerophon proof is inconvenienced by manual definition expansion (\texttt{expandAllDefs}), indexed formula references (e.g., the uninsightful numeric argument $-4$), and manual weaking with \texttt{hideR}.
Moreover, Bellerophon compromises traceability by completely separating its definitions (\texttt{Definitions}), model (\texttt{Problem}), and proof (\texttt{Tactic}).
Kaisar's traceability remains an important readability advantage on every example, independent of the major conciseness advantage seen in the PLDI-TAC example.
The Angelic statements \texttt{for} and \texttt{switch} also arguably improve readability versus Bellerophon's (equivalent) use of duality (\texttt{\textasciicircum@}).

\begin{figure}[!htbp]
\begin{minipage}{\textwidth}  
\begin{lstlisting}[language=kaisar]
let inv() <-> (d>=v*(eps-t) & t>=0 & t<=eps & 0<=v&v<=V);
?(d >= 0 & V >= 0 & eps >= 0 & v=0 & t=0);
for (time := 0; !(inv()); ?(time <= 10000); time := (time + 600);) {
 switch {case (d>=eps*V) => v:=V; ?(0<=v&v<=V); case (true) => v:=0;}
 {t := 0; {d' = -v, t' = 1 & ?(t <= eps) & !(d >= v*(eps-t))};} !(inv());
} !(d >= 0);
\end{lstlisting}
\end{minipage}
~\\
\begin{minipage}{\textwidth} 
\begin{lstlisting}[language=bellerophon]
Lemma "PLDI-TAC". Definitions
  B bounds() <-> (V>0&eps>0).
  B init(R d,R v,R t) <-> (d>=0&bounds()&v=0&t=0).
  B safe(R d) <-> (d>=0).
  B loopinv(R d,R v,R t) <-> (d>=0&t>=0&t<=eps&0<=v&v<=V).
  HP ctrl ::= {{{?d>=V*eps();v:=V;{?0<=v&v<=V;}^@}++{v:=0;}}^@}.
  HP plant ::= {t:=0;{d'=-v,t'=1&t<=eps}}.
End. Problem.
  init(d, v, t) -> 
  [time:=0; {{{?(time<= 10000); ctrl; plant; time:=time + 600;}^@}*}^@
  ] safe(d) End.
Tactic "PLDI-TAC: Proof"
  expandAllDefs; unfold;
  con("k", "(d>=0&t>=0&t<=eps&0<=v&v<=V)&10000-time<=k*600", 1); <(
   auto, auto,
   dualDirectd(1); composeb(1); testb(1); implyR(1); composeb(1);
   dualDirectb(1); choiced(1); orR(1); cut("d>=V*eps|true"); <(
    orL(-4); <(hideR(2); unfold; dC("t>=0", 1); <(
     dC("d>=V*(eps-t)", 1); <(dW(1); auto, dI(1)), dI(1)),
      hideR(1); assignd(1); composeb(1); composeb(1); assignb(1);
       dC("t>=0", 1); <(
       dC("d>=0*(eps-t)", 1); <(dW(1); auto, dI(1)), dI(1))), propClose))
End. End.
\end{lstlisting}
\end{minipage}
\caption{Listings for PLDI-TAC (top: Kaisar, bottom: Bellerophon).
Listings were changed to fit the page.}
\label{fig:pldi-tac-listings}
\end{figure}

\subsection{Case Study Statistics and Discussion}
\label{sec:kaisar-evaluation-stats}
We give statistics for the case studies, which we then discuss in greater detail.
Line counts are given in \rref{tab:kaisar-proof-metrics} and full source listings are given in the extended version ~\cite[App.\ D]{Bohrer21}.
Line counts require careful analysis.  It is encouraging that many Bellerophon proofs got shorter in Kaisar, but further interpretation is needed because short code in a given language does not universally reflect higher productivity. For example, Kaisar consciously chose named assumptions for readability, even at cost of verbosity. Expert users' line counts can also differ from  typical users. To draw deeper conclusions despite the limitations of line counts, we measured the purpose of each line and how much models \emph{change} when maintained.  Even in the presence of syntactic differences, these counts provide insight into the relative effort expended on tasks such as modeling, initial proof attempts, and proof maintenance.

Even then, metrics do not tell the whole story. Kaisar's goals included naming facts, making the textual relationship between model and proof easily traceable, providing a gentle learning curve from systems to games, and providing a gentle learning curve from models to specification and proof. Subjective goals such as these are better appreciated by reading the examples from previous sections, rather than transplanting a subjective motivation onto empirical data.

\begin{table}[!h]
\begin{tabular}{lrrrrll}
Model Name (Bellerophon) & Lines & Model & Proof & Assump & Same + Diff\\\hline
PLDI-DC   & 15   & 13      & 3     & 0     & N/A\\
PLDI-AS    & 42     & 15      & 27    & 9     & 9 + 33 \\ 
PLDI-TAC & 39      & 15      & 24    & 5     & 19 + 20 \\ 
PLDI-RA    & 28     & 19      & 9    & 0     & 10 + 18\\  
PLDI-RAD  & 29      & 20      & 9    & 0     & 27 + 2\\  
IJRR & 88 & 36 & 52 & 3 & N/A\\
RA-L & 294 &  67 & 227 & 97  &  N/A
~\\[4pt]
Model Name (Kaisar)\ \ \ \ \ \ \ \  & Lines & Model & Proof & Assump & Same + Diff\\\hline
PLDI-DC   & 7  & 4   & 3 & 0 & N/A\\
PLDI-AS   & 10 & 7   & 4 & 0 & 6 + 4\\
PLDI-TAC & 9  & 7   & 4 & 0 & 7 + 2\\
PLDI-RA  & 15 & 11 & 6 & 0 & 2 + 13\\
PLDI-RAD & 16 & 12 & 6 & 0 & 11 + 5\\
IJRR     & 62 & 31 & 31 & 12 & N/A\\
RA-L     & 491 &  133  & 372   & 138   & N/A
\end{tabular}

\caption{Proof metrics for Bellerophon proofs and Kaisar ports, respectively.}
\label{tab:kaisar-proof-metrics}
\end{table}
Specifically, we measure lines of model, total lines of proof, and lines of proof which specify the assumptions passed to proof automation.
In \rref{tab:kaisar-proof-metrics}, the Bellerophon results are in the first table and Kaisar in the second.
Sizes are given in non-blank, non-punctuation, non-comment lines.
The total line count can be less than the sum of modeling and proof lines because the same line may contribute to both the model and proof.
Assertions are counted as proof but not model.
The (Assump) column counts all lines which contain \verb|using| clauses in Kaisar and  \verb|hide| (weakening)
steps in Bellerophon. Contrary to Kaisar's positive mention in \verb|using|, the \verb|hide| steps specify that a given assumption should be \emph{unused} by proof automation.
Line counts are based on 70-character lines, except we counted atomic proof steps in the Bellerophon version of RA-L because the line count would be inflated to at least 340 by line-breaking because such verbose proof steps are used. In the difference column (Same + Diff), newly added lines are considered different. 

We describe the example models and proofs displayed in \rref{tab:kaisar-proof-metrics}.
The PLDI-DC~\cite{DBLP:conf/pldi/BohrerTMMP18} model is a 1D, velocity-controlled driving model where the environment (Demon) controls velocities and loop repetition.
The AS, TAC,  RA, and RAD variants respectively extend PLDI-DC with Angelic sandboxing, time-based Angelic loop durations, a reach-avoid analysis, and (Demonic) actuator disturbance.
The IJRR and RA-L driving models have 2D curved motion and acceleration control. IJRR emphasizes its support for inexact sensing and actuation, while RA-L emphasizes relative coordinates, speed limit-following, and inexact waypoint-following.

We discuss proof length. The Kaisar files were shorter, except for RA-L.
One important source of the reduction is that Kaisar removed duplication between models and proofs by combining them into one artifact.  Bellerophon had the shorter RA-L proof because its case analysis rule excels at deduplicating proofs of differential cuts. We plan to provide the same ability, and thus the shorter proof, in Kaisar by allowing \verb|switch| inside a domain constraint. We report this long RA-L proof as a reminder why the evaluation is important: \emph{every} new language will have cases in which it is less elegant than predecessors, but by identifying those cases through practical use, one can often identify simple feature proposals that restore elegance.
The PLDI examples were shorter than RA-L in both languages.
For PLDI-DC, the only Kaisar proof steps were invariants, to which Bellerophon added one invocation of general-purpose proof automation. No manual assumption reasoning was needed. 
As discussed below, the remaining PLDI examples had non-trivial proof scripts in Bellerophon and concise, annotation-style proofs in Kaisar.

In Kaisar, the largest change occurred in PLDI-RA because the transition from a timed paradigm to a reach-avoid paradigm affected almost every part of the model, including assumptions, loop structure, controllers, and invariants.  As PLDI-RAD shows, later changes may affect fewer lines, with PLDI-RAD only changing two lines of model to introduce actuation disturbance and three lines of proof where assumptions on the disturbance are explicitly used. 
In PLDI-RA and PLDI-RAD, the use of multiple line labels (for the initial state and the start of the loop body, respectively) allowed terms and formulas to be written in a stable, maintainable way.
In PLDI-TAC and PLDI-AS in Kaisar, changes were minimal, because their differences in control schemes are expressible in a few lines. In Bellerophon,  the largest change came in PLDI-AS because the proof approach switched from highly-automated proof by annotation to an explicit proof with multiple branches, one for each controller branch. Incidentally, this branching-style proof typically increases the number of \verb|hide|s in Bellerophon. In Bellerophon's defense,
it too achieved nontrivial reuse with auxiliary definitions, but the parts which changed are telling: small conceptual model changes can require many changes in hides, cuts (assertions), and any proof steps which refer to facts by numeric identifiers. These are specific cases which Kaisar sought to, and did, address.
Conversely, Bellerophon's reuse numbers are strongest in proofs (PLDI-RAD) where most assumptions are used in most proof steps. As model size increases, however, proof steps need to minimize their assumptions for performance reasons.
Thus the advantage fades not only because more \verb|hide|s are required, but because maintenance of \verb|hide|s is fundamentally non-local, when compared to maintenance of Kaisar-style assumption lists.

We discuss the IJRR model.
In contrast to RA-L, IJRR had more concise casing structure when expressed in Kaisar.
The concise case structure demonstrated the value of Kaisar's disjunctive lookups: each of the 3 control cases proved a correctness lemma, 
after which a single proof of the ODE is written which automatically appeals to the disjunction of control lemmas.
Notably, the Bellerophon proof had fewer assumption-management lines.
This reflects the fact that the available assumptions were simple enough that automated solvers could prove the assertions with little manual assumption hiding. While Kaisar had more explicit assumption lines, its concise assumption syntax is more readable and maintainable than Bellerophon's, as discussed in \rref{sec:kaisar-introduction}.

We discuss RA-L further. Kaisar's simpler case-analyses led to a longer proof, but RA-L helped stress-test Kaisar with its nested cases and multiple ODEs on different branches. Many lines had \verb|using| clauses: 138. In practice, the user often starts writing assumptions everywhere once they are used anywhere; it is unclear which assumption lines are necessary. Bellerophon had 97 \verb|hide| statements, a smaller percentage difference between languages than in the smaller IJRR example.

We tested Kaisar on all examples in this paper plus a few dozen unit tests.

\section{Conclusion and Future Work}
\label{sec:kaisar-future-work}
This paper presented Kaisar, the first proof language for \CdGL (Constructive Differential Game Logic).
Kaisar is used for proofs of a broad range of correctness properties for a broad class of CPS models: hybrid games.
Because it is important to show that models of CPS behave correctly in \emph{every} scenario, such proofs are also important:
rather than achieving scalability through approximative analyses or restrictions on models or bounds on the correctness of the analyses, scalability is achieved using human insights expressed in the proof.
Proof checkers, including Kaisar, ensure only sound proof rules are used, so that only proofs of valid properties are accepted.
Because writing and maintaining such proofs can require significant time and effort, Kaisar's design emphasizes a
\emph{structured} design aimed  at managing the time and effort required for proofs.
As a particular example, we showed how Kaisar's labeled reasoning feature streamlined support for paradigms including logical model-predictive control, sandboxing, and reach-avoid verification.
Other structured features include block structure, persistent named facts, and definitions.
We argued how these features promote our goals, including readability, maintainability, and traceability, then measured these goals to the extent practical (\rref{sec:kaisar-evaluation}).
The features were supported by novel technical contributions, such as SSA-style variable numbering that  supported high-level reasoning across changing states. In whole, these features provide a smooth learning curve for Kaisar, letting users focus more on developing the key insights of their proofs. The basic design of the Kaisar language and its implementation are likely of broader interest beyond CPS, as well.

\CdGL helps Kaisar make proofs correspond closely to executable code.
A key application is extracting correct implementation code from Kaisar proofs~\cite[Ch.\ 8]{Bohrer21}
by a reduction to strategy synthesis~\cite{DBLP:conf/pldi/BohrerTMMP18} and by the operational semantics of \CdGL~\cite{DBLP:conf/cade/BohrerP20}  strategies.
Games and refinement combine to make synthesis more flexible:
when  refinement shows two strategies play the same game, 
we expect their code is easily swapped
without changing the embedded platform.

Theorem-proving and synthesis, together, show systems correct from design to implementation.

\begin{acks}
We thank the EMSOFT reviewers for their feedback.
This research was funded by the Alexander von Humboldt Foundation, the NDSEG Fellowship, the Siebel Scholarship, and by the AFOSR under grant number FA9550-16-1-0288.
\end{acks}

\bibliographystyle{splncs04}
\bibliography{kaisar,platzer}

\end{document}